# Resource Orchestration and Optimization in 6G Extreme-edge Scenario

Manuel A. Jimenez [1], Sarang Kahvazadeh[2], Ignacio Labrador [1], Josep Mangues-Bafalluy[2]

[1]EVIDEN, [2]Centre Tecnologic Telecomunicacions Catalunya (CTTC)

[1]MADRID, [2]Castelldefels, Spain.
Email: [1]manuel.jimenez@eviden.com [2]skahvazadeh@cttc.es cttc.es [1]ignacio.labrador@eviden.com [2]josep.mangues@cttc.es

*Abstract*— 6G networks envision a pervasive service infrastructure spanning from centralized cloud to distributed edge and highly dynamic extreme-edge domains. This vision introduces significant challenges in orchestrating services over heterogeneous, volatile, and often mobile resources beyond traditional operator control. To address these challenges, this demo presents a 6G-ready orchestration architecture focused on resource prediction and service resilience at the extreme-edge. The proposed solution integrates (i) an AI/ML-based Infrastructure Status Prediction Module, (ii) a Monitoring System capable of handling large-scale, diverse telemetry, and (iii) a Decision Engine and Actuator that ensures proactive service migration and continuity in unstable environments.

## I. Introduction

One of the primary research challenges in achieving effective service management and orchestration for 6G is known as continuum orchestration. This concept involves the seamless control of network functions and resources across cloud, edge, and emerging extreme-edge domains, treating them as a single, borderless infrastructure. In a 6G environment, this continuum will extend beyond the resources owned by a communications service provider (CSP) to encompass a vast and inherently volatile array of devices. These devices can be found in factories, vehicles, homes, streets, railways, satellites, and smart city installations, in addition to users' personal terminals.

Extreme-edge devices are already emerging through mobile edge computing (MEC) and large-scale IoT deployments. 6G will amplify this trend, realizing the long-held vision of pervasive networks in the mobile domain. Unlike previous generations of handsets, many next-generation devices will have significant computing and storage capabilities. As a result, they should be viewed not only as sources or sinks of traffic but also as additional infrastructure nodes capable of hosting microservices close to the user. This approach reduces latency and expands the CSP's effective footprint, particularly in underserved rural or disaster-affected areas.

However, taking advantage of this opportunity is challenging. Extreme-edge resources are often located beyond the operator's direct control, subject to intermittent connectivity, and varying widely in capabilities. Additionally, these resources may also be mobile. This diversity and dynamism require a new class of orchestration strategies that can predict resource availability and respond—or ideally, proactively adapt—to sudden changes in network topology. Also resource optimization in case of Extreme-edge is one of the remain challenges.

To overcome Extreme-edge resource orchestration and optimization, we propose an architecture as demonstration including:

- Analytic Engine (AE): This component is named Infrastructure Status Prediction Module (ISPM) which rely on AI/ML techniques to predict the activation status of various infrastructure devices across multiple network slices.

- Monitoring System (MS): The monitoring system shall be enabled with the mechanisms to collect and process monitoring and diagnostics data from a big amount and diversity of devices in the different network domains, including the extreme-edge.

- Decision Engine (DE) and Actuator: It is a central component responsible for intelligent orchestration and automated migration of services. Its primary function is to ensure service continuity and optimal resource utilization by actively monitoring the health and status of compute nodes, and dynamically re-assigning services as the environment changes in Extreme-edge.

## II. Proposed Demo Architecture

### A. The Architecture

The proposed architecture is illustrated in Figure 1. In the figure:
- Infrastructure Layer Emulator is an simulation of Extreme-edge.
- MS responsible for collecting and processing the data received from the network infrastructure. This

component also requests the prediction to the AE and forward it to the Decision Engine (DE).

- ISPM is and AI/ML component acting as AE. This component rely on AI/ML techniques to predict the activation status of various infrastructure devices across multiple network slices. The historical device connection and disconnection information would serve as input for training the ISPM AI/ML predictive models, which in turn would be able to forecast the connectivity status of each device in the infrastructure. The viability of this approach is supported on the demonstrated effectiveness of specific AI/ML algorithms addressing tasks regarding time series processing [1], and also, handling substantial volumes of data [2] (essential for managing data across a multitude of slices), and their capability to correlate information from diverse sources to discern behavioral patterns that might elude human perception [3] [4]. The ISPM predict the node´s ON/OFF.
- The DE and ACT in this demo is implemented as one component. The DE getting input from ISPM periodically ON/OFF nodes status then consider the CPU/Memory of ON nodes and make ACT properly to migrate services without service corruption.

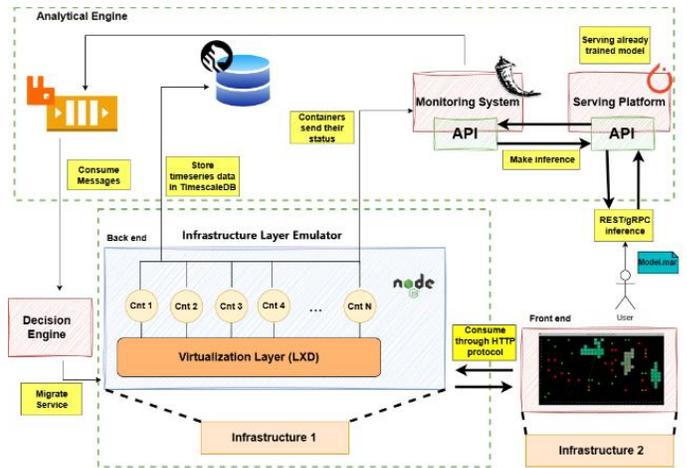

Figure 2. Demonstration technological components

**Infrastructure Layer Emulator:** The development of the ILE is being addressed considering two main components: its backend (which is the actual emulator) and a front-end (a GUI connected to the backend and representing in near real time the changes in the infrastructure layer, as well as the services deployed on it). The backend is being developed relying on the LXD technology and a set of ancillary shell scripts, while the front-end is being developed as a Java application.

**Analytic Engine**: The Analytical Engine (AE) implements a time series prediction model using a Long Short-Term Memory (LSTM) network in Pytorch. Its primary purpose is to predict the future status of a set of containers based on historical binary status data and timestamp information. The model is particularly well-suited for sequential data and is designed to learn temporal patterns in how the containers behave throughout the day. It is designed to generalize across multiple containers and time intervals, and its output is a multivariate binary prediction indicating the expected state of each container at a future point in time, which is the input of the DE to take decision according to this predictions and others variables. The data for training the model is loaded from a TimescaleDB, an SQL database optimised for time-series. The input dataset for training includes the binary status of each container over time and a "ground_truth" value, along with a timestamp. The continuous input is derived from the timestamp — specifically, the time of day is normalized and transformed using sine and cosine functions to represent its cyclic nature. This encoding helps the LSTM network understand patterns related to daily cycles. Ground truth label represents the "answer key" for training, it means the true state that the model should predict, that are also retrieved from the database. Finally, the AE's model is served using TorchServe, which is a flexible and production-ready model serving library developed by PyTorch. In particular, TorchServe is deployed using official docker image aligned with cloud-native requirements. There are two phase of AE's workflow:

- In the training phase, the user launches the training and data requests are made to the database, this database contains the state of each device at each moment, and the "answer key" called "ground_truth". The preprocessing module normalizes the timestamps and sends these

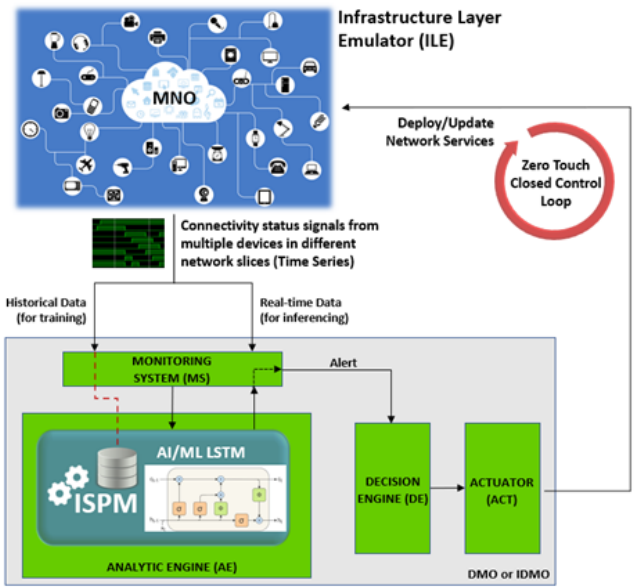

Figure 1. Demonstration Architecture

B. ARCITECTURE COMPONENTS DESCRIPTION

Figure 2 illustrates technological architecture of demonstration in detail:

- sequences to the trainer, which performs the training cyclically and saves the model.
- In the service phase, the administrator serves the model (saved in the previous phase) to the AI server (Torch Server), which performs its processing, after that the MS can make inferences through the API.

**Monitoring system:** The Monitoring System (MS) is responsible for collecting and processing the data received from the network infrastructure. This component also requests the prediction to the AE and forward it to the DE. All these interactions and interfaces are implemented using API Rest that offers high scalability, flexibility and is a widely used and well-established HTTP. The API Rest used by MS is implemented using flask. All the collected data are sent to the RabbitMQ microservice.

**Discovery of each nodes:** The system updates the status of each node through an API Rest which collects data from all nodes and processes it to be served in the correct format to the AE. The endpoint it's called "report-state", and it is implemented through a POST action using Flask library. For this demo, the nodes report the following information to the MS: TimeStamp, indicates the time when the container sends the update. State that, in this particular case, it is considered only two states (ON/OFF). Container ID, an identifier that identifies each container univocally and , it is assigned at the deployment phase. From the point of view of the MS, that process is a reactive mechanism, since it is each node that starts the process of the discovery.

**Decision Engine and Actuator:** DE continuously listens to updates regarding the operational status of all nodes in the Kubernetes cluster, which are communicated via a message bus (RabbitMQ). Each node may transition between ON (available) and OFF (unavailable) states, for example due to failures, scheduled maintenance, or network disruptions. The DE ensures that key services are always running on available (ON) nodes and are moved promptly from nodes that become unavailable. Rather than statically assigning services to nodes, the DE considers the real-time resource capacity (CPU and memory) of each available node. For every service, it calculates a dynamic "score" for each ON node, based on their current allocatable CPU and memory resources, giving preference to nodes with higher available capacity. This prevents overloading of nodes and contributes to the overall performance and resilience of the cluster. If a node hosting a service is predicted to go offline, or a better resource becomes available, the DE triggers a migration. This involves temporarily marking the current node as unavailable for new workloads (via "tainting") and reallocating the service to a new node, ensuring minimal disruption.

To implement communication between the MS and the DE, RabbitMQ has been used, which is an open-source message broker that delivers messages to a receiver (DE) in an asynchronous way. This characteristic makes it great for our purpose: decoupling services (easier to integrate components from different stakeholders) and scaling systems. It is also a cloud native option due to the fact that there are official docker images and is oriented to microservices communication. Regarding the data of the system, TimescaleDB has been used under the same assumptions as RabbitMQ: Scalability and feasibility to integrate and cloud-native orientation. It is also a time-series database to manage large volumes of time-stamped data, which is an important feature to manage data from the containers and infrastructure monitoring. There is also a Graphic User Interface that has been used to show the migrations applied by the proposed system during the simulation of the volatile infrastructure.

A short conceptual version of demo description can be found at [5].

## III. CONCLUSION

This demo showcases a practical implementation of continuum resource orchestration and optimization tailored for 6G scenarios, with a focus on addressing the unpredictability of extreme-edge environments. By leveraging AI/ML-driven infrastructure prediction, real-time monitoring, and intelligent decision-making, the system ensures resilient service management across a heterogeneous and dynamic infrastructure. The demonstrated architecture paves the way for future 6G deployments that require robust, proactive, and distributed orchestration beyond the traditional network edge.

ACKNOWLEDGMENT

This project has received funding from the National Spanish MINECO under grant No. TSI-063000-2021-54 (6GDAWN-ELASTIC), grant No. PID2021-126431OB-I00 funded by MCIN/AEI/ 10.13039/ 501100011033 (ANEMONE).